\documentclass[preprint]{vgtc}               

\ifpdf
  \pdfoutput=1\relax                   
  \usepackage{graphicx}                
  \DeclareGraphicsExtensions{.pdf,.png,.jpg,.jpeg} 
\else
  \usepackage{graphicx}                
  \DeclareGraphicsExtensions{.eps}     
\fi%

\graphicspath{{figures/}{pictures/}{images/}{./}} 

\usepackage{microtype}                 
\PassOptionsToPackage{warn}{textcomp}  
\usepackage{textcomp}                  
\usepackage{mathptmx}                  
\usepackage{times}                     
\usepackage{cite}                      
\usepackage{tabu}                      
\usepackage{booktabs}                  

\usepackage{soul} 
\usepackage{balance}
\usepackage{xcolor,colortbl} 
\usepackage[hyphens]{url}
\usepackage{amssymb} 

\onlineid{1515}

\vgtccategory{Research}

\vgtcinsertpkg

\title{Working with XR in Public: Effects on Users and Bystanders}

\author{
 Verena Biener$^{1}$\thanks{e-mail: verena.biener@hs-coburg.de}
\and Snehanjali Kalamkar$^{1}$
\and John J. Dudley$^{2}$
\and Jinghui Hu$^{2}$
\and Per Ola Kristensson$^{2}$
\and Jörg Müller$^{3}$
\and Jens Grubert$^{1}$\thanks{e-mail: jens.grubert@hs-coburg.de}
}
\affiliation{\scriptsize $^{1}$Coburg University of Applied Sciences and Arts, Germany\\  $^{2}$University of Cambridge, United Kingdom \\  $^{3}$University of Bayreuth, Germany}

\abstract{
Recent commercial off-the-shelf virtual and augmented reality devices have been promoted as tools for knowledge work and research findings show how this kind of work can benefit from the affordances of extended reality (XR).
One major advantage that XR can provide is the enlarged display space that can be used to display virtual screens which is a feature already readily available in many commercial devices.
This could be especially helpful in mobile contexts, in which users might not have access to their optimal physical work setup. Such situations often occur in a public setting, for example when working on a train while traveling to a business meeting. At the same time, the use of XR devices is still uncommon in public, which might impact both users and bystanders. Hence, there is a need to better understand the implications of using XR devices for work in public both on the user itself, as well as on bystanders.
We report the results of a study in a university cafeteria in which participants used three different systems. In one setup they only used a laptop with a single screen, in a second setup, they combined the laptop with an optical see-through AR headset, and in the third, they combined the laptop with an immersive VR headset. In addition, we also collected 231 responses from bystanders through a questionnaire.
The combined results indicate that (1) users feel safer if they can see their physical surroundings; (2) current use of XR in public makes users stand out; and (3) prior XR experience can influence how users feel when using XR in public.

} 


\teaser{
  \centering
  \includegraphics[width= 0.95\linewidth]{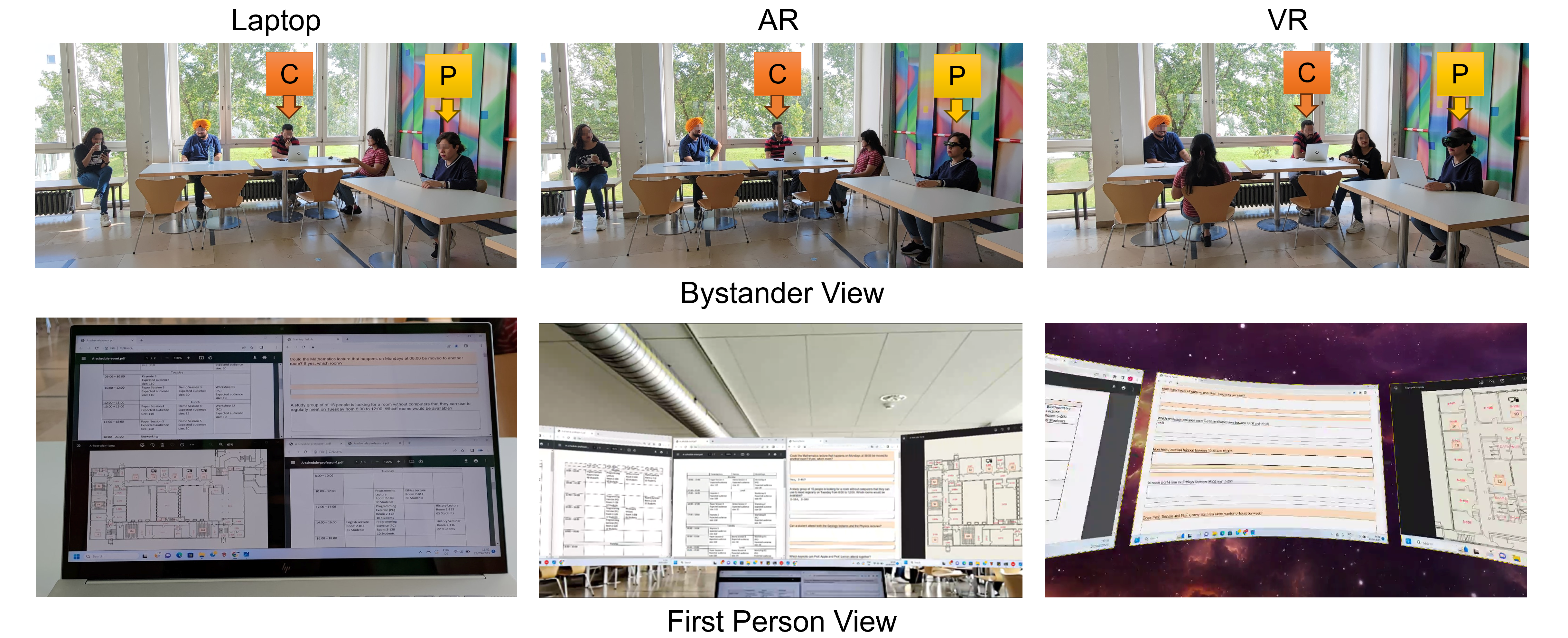}
  \caption{The top row shows the view of a bystander passing the area of the study for all three conditions (\textsc{laptop}, \textsc{ar}, \textsc{vr}). C indicates the coordinator of the study and P indicates the participant. The illustrations are reenactments of the actual situations. The bottom row shows the participant's views during the task for each condition. }
  \label{fig:teaser}
  \vspace{-0.2cm}
}

\begin{document}


\firstsection{Introduction}

\maketitle

Using extended reality (XR) for knowledge work has recently gained popularity through the promotion of current commercial off-the-shelf devices as tools for work, such as the Meta Quest Pro and the Lenovo Think Reality glasses. In addition, prior research has also demonstrated the potential benefits of using XR for knowledge work, such as how virtual reality (VR) can be used in open office spaces to reduce distractions \cite{ruvimova2020transport} and how new interaction possibilities provided by VR can improve the performance for certain tasks \cite{gesslein2020pen,biener2022povrpoint}. In addition, researchers have evaluated how the large display space provided by augmented reality (AR) and VR can be used in the context of knowledge work \cite{mcgill2020expanding, pavanatto2021we}.
In theory, the increased display space could also increase productivity \cite{czerwinski2003toward} and is already available in current commercial head-mounted displays (HMDs).

These properties of XR seem especially advantageous in mobile scenarios, where the work environments are less optimal, such as in crowded spaces with many distractions, or in confined spaces that limit the size of physical hardware \cite{grubert2018office}, for example, in public transportation or at a coffee shop.
Prior studies on supporting knowledge work have focused on new technologies evaluated mainly through laboratory-based experiments. These are typically not representative of public spaces,  with often only those people involved in the study present.
However, the aforementioned scenarios that could benefit greatly from using XR, are in fact frequently of a semi-public or public nature. Therefore, it is interesting to explore how users of XR devices \textit{feel} while working in public spaces, and also to investigate what people around them think and how they react to the currently relatively rare sight of XR-users in public.

However, so far, this research question has not been explored sufficiently. To this end, we report the results of a study in which participants completed three sessions of working in a public space, once with a standard laptop,  once with a laptop combined with an optical see-through AR headset, and once with a laptop combined with an immersive VR headset.
The results gathered through questionnaires, interviews and observations provide detailed insights into how using XR in public spaces is perceived today. Specifically, our results indicate that users can feel safer when they are able to see their physical environment, that XR still imposes more physical burdens on the user, has lower usability than a laptop (about 7\% for AR and 19\% for VR), as well as that previous XR-experience and personality can influence how the use of XR-devices in public is perceived. We also found that using XR in public currently makes users stand out and that many bystanders were curious about the XR-device, yet they had no clear understanding of what those devices are used for.



\section{Related Work}

\subsection{Extended Reality for Knowledge Work}
\label{sec:MixedRealityForKnowledgeWork}

Research has indicated that using XR for knowledge work can potentially provide benefits to users. 
One apparent benefit of XR could be its large display space, which has been studied in various contexts \cite{mcgill2020expanding, ens2014personal, pavanatto2021we, ng2021passenger, medeiros2022fromshielding} and has also recently been promoted by XR solution providers such as Apple \cite{applevisionpro}, Lenovo \cite{lenovothinkreality}, Meta \cite{metaquestpro}, or Sightful \cite{sightful}, as a tool for supporting knowledge work.
Ens et al.~\cite{ens2014personal} showed that task-optimizing layouts can greatly improve task switching.
On the other hand, Ng et al.~\cite{ng2021passenger} found that in a mobile scenario, other passengers' presence can influence the layout preferences of the user, which also relates to the findings of Medeiros et al.~\cite{medeiros2022fromshielding} who detected that users try to avoid placing virtual displays above other passengers.
Pavanatto et al.~\cite{pavanatto2021we} have studied the use of virtual monitors with regard to several aspects such as performance, accuracy, and comfort. They found that this can be beneficial, but are still inferior to physical screens, and, therefore, suggest combining virtual and physical monitors.

Knowledge work can also benefit from multimodal interaction possibilities provided by XR, including eye-tracking, gestures, spatially tracked devices and three-dimensional visualizations.
Biener et al.~\cite{Biener2020Breaking} used a combination of touch and eye-tracking to more efficiently interact with multiple virtual screens. McGill et al.~\cite{mcgill2020expanding} altered the position of virtual displays depending on the user's gaze-direction to decrease head movement. VR has been combined with a touch-screen and a spatially tracked pen in the context of, authoring presentations \cite{biener2022povrpoint} and spreadsheet applications \cite{gesslein2020pen}, which also benefits from the three-dimensional display space. In addition, XR can enhance the capabilities of standard devices such as keyboards \cite{schneider2019reconviguration}, or address privacy issues, especially in a mobile context ~\cite{grubert2018office, schneider2019reconviguration} as the virtual content is not visible to bystanders.

Further, VR can be used to reduce distractions, for example in an open office environment \cite{ruvimova2020transport}. Also, AR has been shown to help against distractions \cite{lee2019partitioning} and can be used, as well as VR, to personalize the work environment, which is often not possible in shared workspaces. However, Li et al.~\cite{li2021rear} found that even though users preferred natural environments, they were more productive in office-like environments.

Through the possibility to exchange the physical surroundings with a virtual one, XR also has the potential to reduce stress, for example, by experiencing a virtual nature environment  \cite{anderson2017relaxation, thoondee2017using, thoondee2017usingvirtual} and it has been shown that this works better when using VR than watching a video on a normal display \cite{pretsch2021improving}. Also, an active VR experience can have more positive effects on stress than a passive one \cite{valtchanov2010restorative}. 

We have seen that XR can provide benefits by shielding the users from the physical world. However, this can also be seen as a drawback, as they might not have access to relevant parts of the physical world. In addition, presence can be improved when integrating real-world distractions \cite{tao2022integrating}. Therefore, McGill et al.~\cite{mcgill2015dose} suggested integrating such parts into the virtual environment.  For this purpose, Hartmann et al.~\cite{hartmann2019realitycheck} proposed to include real-time 3D reconstructions of the physical surroundings into VR and Wang et al.~\cite{wang2022realitylens} presented a physical world view that can be customized. Different methods have also been proposed to include smartphones into the virtual environment \cite{alaee2018user, desai2017window, bai2021bringing}.
Displaying bystanders in the virtual world can be especially important, as O'Hagen et al.~\cite{o2020reality} found that users feel uncomfortable not knowing the position of bystanders which can, for example, be detected and displayed using depth cameras  \cite{simeone2016vr}, but such a functionality is not yet readily available in many commercial devices.
By studying the use of off-the-shelf VR and AR devices in a public space, we contribute a better understanding of how participants use these devices and how they cope with their limitations, such as not being fully aware of the physical environment.

Another drawback of using XR for knowledge work could be the form factor of current devices. A recent review \cite{souchet2022narrative} showed that using VR can lead to increased stress, mental overload, visual fatigue, and muscle fatigue. VR has also been found to result in higher physical discomfort than a standard physical work environment \cite{guo2020exploring}. Biener et al.~\cite{biener2022quantifying} found that VR resulted in worse ratings than a physical environment for a range of measures such as task load, usability, flow, frustration and visual fatigue.
An in-the-wild study by Lu et al.~\cite{lu2021evaluating} pointed out that participants could imagine using the AR system daily, if the form factor of the devices would be improved.

The study most similar to ours was presented by Li et al.~\cite{li2022designing} who compared a tablet with an AR and VR HMD during a typing task in the backseat of a car. They found that by using the HMDs the users had a lower awareness of their real surroundings, increased perceived workload, and lower text entry performance as compared to the tablet condition. However, through a customized workspace layout participants could improve their perceived performance and decrease head movements when switching between keyboard and display.  Our work also compares AR and VR to a standard device, in our case a laptop, in a mobile setting. However, our focus lies on evaluating the effects of using such systems in a public environment.

\subsection{Social Acceptability}
Through the large display space and the novel interaction possibilities, XR could be especially useful in mobile contexts with limited space or the limitation of using small portable devices.
However, this opens up the question of social acceptability. How do users feel when using such devices in spaces that are more public than their homes or offices and also, how do other people around them react?

There are various definitions of social acceptability. 
Schwind et al.~\cite{schwind2018need} proposed a definition stating that higher social acceptability makes others more interested in including the user of the technology in their social groups. This suggests that others do not have strong negative feelings towards the user, which is also how we understand the term social acceptability in our paper.
Koelle et al.~\cite{koelle2020social} stated that an interface is socially acceptable if it is “consistent with the user's self-image and external image, or alter them in a positive way”. They also suggested design strategies to enhance social acceptability such as being subtle, designing devices to look like non-digital accessories and giving the bystanders some idea of what the user is doing. 
Gimhae \cite{gimhae2013six} identified six factors that play a role in the acceptance of wearable computers which include how they affect social interactions, the fulfillment of basic needs, and the wearer physically, but also the technical experience, demographic characteristics and whether the device is useful plays a role.  
Koelle et al.~\cite{koelle2020social} also criticize current research which often asks participants to imagine situations instead of letting them experience them.
Yet, Alallah et al.~\cite{alallah2018crowdsourcing} compared data on the social acceptability of HMDs obtained through online surveys with data obtained through a study conducted in a busy public place and found no significant differences.
Vergari et al.~\cite{vergari2021influence} simulated different social environments with 360° videos to evaluate the effects of using VR in these environments on user experience and social acceptability.
Schwind et al.~\cite{schwind2018virtual} evaluated the social acceptance of VR glasses through an online study, showing participants images of VR users in different situations and found that it is less acceptable in social situations, such as in a living room or a cafe.
Similarly, Profita et al.~\cite{profita2016effect} showed videos of VR use in public and found that it was more socially acceptable, if the head-mounted display was used as a tool to support persons with a disability.

However, there have also been studies in which participants could actually use a VR device, such as by George et al.~\cite{george2019should} who explored in a lab study if bystanders can identify task switches of VR users and which strategies they employ to interrupt them.
Other studies have actually been conducted in a public place.
For example, to evaluate the use of interactive glasses showing notifications while walking through a calm and busy street \cite{lucero2014notifeye}.
Another study conducted in a Starbucks cafe, explored different interaction techniques and form factors of smart glasses \cite{tung2015user} and Lee et al.~\cite{lee2018designing} presented an elicitation study for socially acceptable hand-to-face input conducted in a busy public space.

Also, through a field-experiment, Eghbali et al.~\cite{eghbali2019social} found that VR users' experience is influenced by factors such as freedom to switch between realities, sense of safety, physical privacy and uninterruptible immersion while bystanders' experience depends on feeling normal, sense of safety and shared experience.
As our goal was to gain direct insights into how users and bystanders feel about using XR in public, we decided to conduct our study in a public space where participants can experience first-hand how they \textit{feel} while using AR and VR HMDs and which also allows us to get direct responses from bystanders.
In the context of knowledge work, this is still under-explored.

To measure social acceptability, different questionnaires and scales have been used, mainly focusing on how bystanders feel when watching someone use a certain device.
Kelly et al.~\cite{kelly2016wear} proposed the WEAR scale measuring the social acceptability of wearable devices using a set of 50 questions and through a study found that social acceptability can be rated differently by varying groups \cite{kelly2018wearer}. Further research extends on this questionnaire \cite{schwind2020anticipated} or uses different questions asking participants to rate their experience while watching someone use a certain technology \cite{schwind2018virtual, meyer2019scenario, kelly2016phd, profita2016effect}.
Ahlström et al.~\cite{ahlstrom2014you} focused on the perception of the user asking them to rate their overall impression during the task and asking them where and in front of whom they would feel comfortable performing the proposed gestures.
Koelle et al.~\cite{koelle2015don} asked participants to rate their feelings while using the device on pairs of opposites such as threatened and safe or unsure and self-confident.
Similarly, Verma et al.~\cite{verma2015study} asked participants to rate their feelings on Likert scales ranging from embarrassed to comfortable and from foolish to sensible.
In our study, we used a subset of these questionnaires to obtain a measure that describes the feelings of our participants and the bystanders.

\section{User Study}
Our goal was to gain insights into how different measures such as user experience, well-being, and productivity in a public space are influenced by different interfaces including a standard laptop, an AR device and a VR device.
The design, apparatus and procedure described in the following were informed by a pilot study with three participants in which we tested varying parameters such as screen resolution and the exact study-procedure.


\subsection{Study Design}
The study was conducted as a within-subjects design with one independent variable \textsc{interface} having three levels, \textsc{laptop}, \textsc{ar} and \textsc{vr}.
\textsc{laptop} was chosen as the baseline, as it is common to use a laptop for working in public spaces.
For \textsc{ar} and \textsc{vr}, we chose commercially off-the-shelf headsets with current software that can connect to a laptop as an input device while displaying multiple virtual screens.
Specifically, we used the Meta Quest Pro in combination with the app “Immersed” \cite{immersed} for \textsc{vr} and the Lenovo Think Reality glasses A3 with its built-in display manager. Both devices were designed and promoted as a tool for knowledge work.

As dependent variables, we captured the perceived task load measured using the NASA TLX questionnaire~\cite{hart1988development}, the usability of using each device through the system usability scale~\cite{brooke1996sus} and a measure of the flow experience~\cite{rheinberg2003erfassung}.
Further, we decided to get an indication of participants' well-being through the simulator sickness questionnaire~\cite{kennedy1993simulator} and by capturing visual fatigue using six questions as done in \cite{benedetto2013readers}.
To gain some insights into participants' emotional state, 
we  asked them, “What was your overall impression / emotion during the task?” on a scale from 1 (I hated it, terribly awkward) to 6 (I enjoyed it, it felt comfortable), c.f. \cite{ahlstrom2014you}. 
In addition, we employed five opposite word pairs as presented by Koelle et al. \cite{koelle2015don} such that participants indicated their feelings on an 11-point Likert scale with the extremes being labeled with these word pairs: tense - serene, threatened - safe, unsure - self-confident, observed - unobserved, skeptic - outgoing. We counted the number of finished tasks in each condition to quantify productivity.

In addition, we conducted a semi-structured interview after each condition about the participants' behavior, what they liked and disliked, in which locations and in front of what kind of people they would be comfortable using such a system and if they would like to use it in the future.
After all conditions, we also asked about their overall preference and with which device they felt most productive and most comfortable.
Throughout the study, we tracked the active window. 

In each session, we asked bystanders to fill out a short online questionnaire including demographic questions, questions about their technical affinity \cite{karrer2009technikaffinitat}, and their experience with working in public and with XR devices.
In addition, they were asked whether they noticed the participant, what their feelings were about this (similar to \cite{verma2015study}), and what they thought the participant was doing. They were also asked to generate an anonymous code, made up by concatenating certain characters from their personal information, which later allowed us to detect multiple answers from one person.

\subsection{Task}
To have a quantifiable measure of productivity, we defined a task in which the progress could be measured unambiguously.
Therefore, we came up with the scenario of planning university class schedules and a conference which should happen in the same building as the classes.
Participants were provided with two PDFs each containing the class schedules of four professors and an additional PDF with the proposed schedule for the conference. In addition, participants got three floor plans one for each level of the building in three separate images.
The class and conference schedules contain information on the number of expected attendees for each class or session, as well as if a computer is needed. The class schedules also show in which rooms the classes are generally held. The floor plans display the room numbers as well as a number indicating how many people fit in the room and a computer icon, if the room provides computers.
Similar to the study conducted by Pavanatto et al. \cite{pavanatto2021we} the participants got a range of questions that they needed to answer by consulting the available documents.
All of these questions required the participants to view at least two different documents and compare their information to give the correct answer, taking into account the room size, number of attendees, if a computer is needed and if there are relevant overlaps.
An example of a task would be to figure out if a specific lecture can be moved to a different room. Therefore, the user needs to check the requirements for that room in the schedule, find rooms in the floor plan that match these requirements and then confirm again with the schedules that the room is not already booked at that specific time-slot.


To reduce the learning effect across the three conditions, we created 3 different sets of questions and documents.
To limit the influence of varying task difficulty, we created them to be comparable, by using the same number of usable Rooms (30, including 10 computer-rooms), the same total number of classes (90), and conference sessions (40). 
Also, the questions were very similar with similar answer lengths and they were presented to the participants in the same order, as the time needed to answer a question could vary.


\subsection{Apparatus}
For the \textsc{laptop} condition we used an HP Envy laptop with a 16-inch display and an external mouse. The laptop's resolution was kept as recommended in the device, 2560x1600 at 150\%.
The same laptop was also used for \textsc{vr} and \textsc{ar}. As the coloring of the keys was hard to read (white on silver), we attached stickers to the keyboard with black font on a white background to make the keys more readable in \textsc{ar} and especially in \textsc{vr}.

For \textsc{ar}, we combined the laptop with Lenovo Think Reality A3 smart glasses, which were connected to the laptop with a USB-C cable. Upon connection, Lenovo's “Virtual Display Manager 3.0.30” \cite{lenovodisplaymanager} started which allowed the user to add and arrange multiple displays.
For AR, we set the resolution to 1920x1080 at 125\% which was found to be most comfortable through internal testing.

\begin{figure}[t]
	\centering 
	\includegraphics[width=1.0\linewidth]{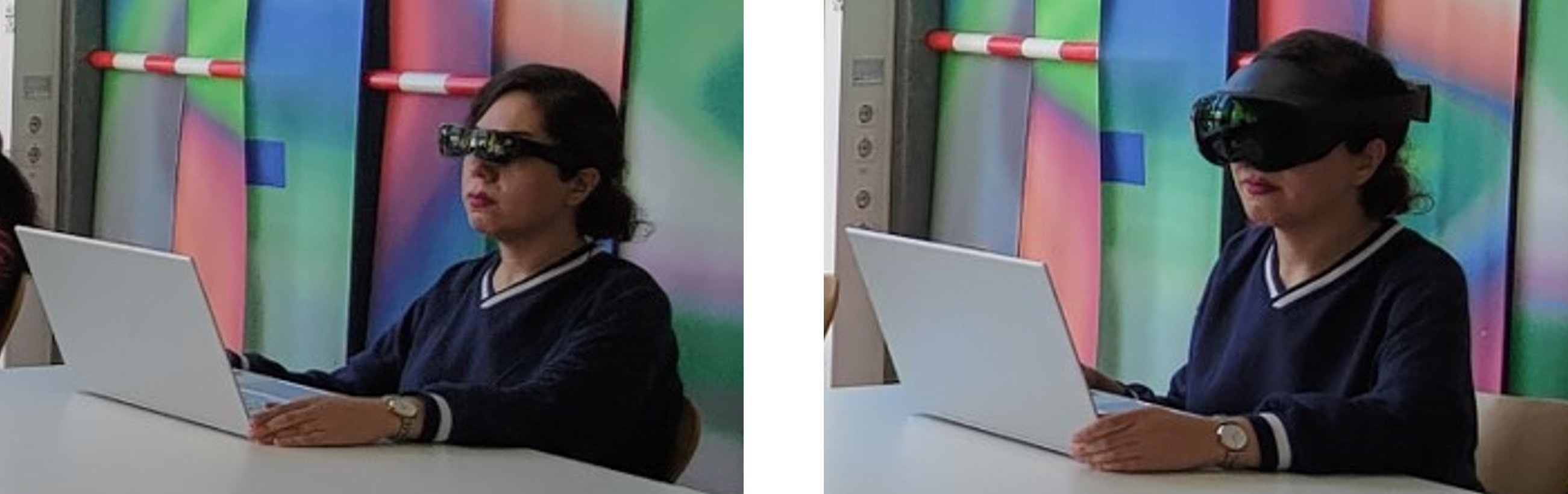}
	\caption{Left: Participant during \textsc{ar} condition wearing the Lenovo Think Reality A3 smart glasses. Right: Participant during \textsc{vr} condition wearing the Meta Quest Pro.}
	\label{fig:close-up}
 \vspace{-0.3cm}
\end{figure}

For \textsc{vr}, a Meta Quest Pro was used in combination with the app “Immersed” \cite{immersed}, which enabled a wireless connection to the laptop and then streamed the laptop content to the virtual screens. Immersed then allowed participants to position the screens, choose one of nine available virtual environments, their preferred seat as well as enable a pass-through window to see the laptop keyboard, mouse, and wherever they felt like they wanted to see the real world. To increase the comfort of wearing the Quest Pro, we added an additional head-strap which distributes the weight more on the center of the head.
In both \textsc{vr} and \textsc{ar}, we set the number of screens to three, which could be viewed comfortably without moving the whole body and allowed distributing all the required files in a way that made them visible at the same time. Through internal testing, these settings were found to be most optimal for each condition using off-the-shelf technologies.
Both setups are depicted in Fig. \ref{fig:close-up}.

\subsection{Procedure}
\label{sec:userstudyprocedure}
Before starting the study, participants were asked to sign a consent form, answer demographic questions, the HEXACO questionnaire that captures certain aspects of their personality \cite{lee2004psychometric} and the technical affinity questionnaire (TA-EG) \cite{karrer2009technikaffinitat}. 

The study was then conducted in three sessions, one for each condition.
These sessions were scheduled at the participants' convenience, yet always happened at a similar time of day to keep outdoor factors as similar as possible. On average, the difference between the earliest and latest start time in the public space was 57 minutes. The largest time difference was 2 hours 23 minutes. We counted the number of bystanders at the beginning and end of each session and calculated the average from this, which was 19 bystanders ($sd=7$) with a minimum of 7 and a maximum of 37. The study was conducted by two coordinators who conducted two sessions together and otherwise, each coordinator conducted around half of the sessions.

All sessions started in a lab at the university. In the first session, the procedure and the study task were explained to the participants. In the following sessions, they were shortly reminded.
In \textsc{vr} and \textsc{ar} they were introduced to the functionality of the devices and taught how to set up the work environment on their own and to their preferences. This included plugging in or connecting the devices, setting the IPD and adjusting the virtual monitors. In \textsc{vr} participants were taught how to choose a virtual environment and how to add a pass-through window for seeing the keyboard and additional passthrough windows if liked.
Then, the participants started the training to get familiar with the task and the device. The training was done in two parts, the first part of the training stopped automatically after 5 minutes to give participants the chance to ask questions or change the setup and then the second part of the training continued for another 15 minutes.
Then, they were reminded of what to do in the cafeteria before the coordinators and participants went there.

In the cafeteria, the participants were assigned a table by the coordinator, which was always on the same side, if possible, with the participants facing toward the bystanders.
The participants sat down and started to set up the work environment.
Meanwhile, the coordinator sat down somewhere else to get a good view of the whole room.
Participants and the coordinator could chat on Discord to clarify any issues. After everything was set up, the participants again started with a short 5-minute task after which they had the opportunity to adjust the setup. The main task then lasted 30 minutes in which participants were asked to answer as many questions as possible and as accurately and completely as possible.
Right afterwards, the participants were asked to answer the questionnaires, while still wearing the HMDs in \textsc{vr} and \textsc{ar}.

During the whole time, the coordinator was taking notes on the events in the room, observing both the participants and other people.
After the participants finished with the questionnaires, the coordinator asked several people in the room to answer a short online-questionnaire accessible by scanning a QR-code.
This was done after the participants finished with the questionnaires so they would not be influenced in their answers in case they noticed people suddenly talking about or staring at them, as this would not be natural but induced by the coordinator.

The bystanders were selected by the coordinator to be representative of the whole room, by choosing people close to the participant, people who had a good view of the participant, and people who sat further away from the participant.
On average, four bystanders were asked to fill out the questionnaire in each condition. Approximately 43\% were sitting close to the participant (2 tables to left and to right, or on the opposite sides including two tables to left and right). As an incentive for filling out the questionnaire, they were offered candies.

Thereafter, the participants and coordinator went back to the private room to conduct the interview. The overall process lasted on average two hours per session. Participants could either count this as their work time or get a gift card.

Due to some technical difficulties, 3 participants had to restart the main task after 3 minutes (P24 in AR), 10 minutes (P22 in Laptop), and 15 minutes (P14 in Laptop). However, we did not exclude them as the learning effect would only be limited, due to the complexity of the answers.
One participant closed the questionnaire window by mistake, less than one minute before the end.

\subsection{Participants}
Eighteen participants successfully completed the study (5 female). Their mean age was 25.67 years ($sd=5.2$). They were all students or employees at our university. All but one have studied or worked in a public space before, generally using a laptop. 
Seven had no prior experience with \textsc{ar}, 9 had slight experience, one person had moderate and one person had much experience. No one ever used \textsc{ar} in public before.
With \textsc{vr}, 10 people had slight experience, 4 moderate, 2 much and 2 very much experience.
Only one person has used \textsc{vr} in public before which was in the university library.
They rated the frequency of using a multi-monitor setup on a scale from 1-never to 5-always. Three participants each rated their use a 1, 2 and 3, six participants rated it 4 and 3 participants rated it 5. Therefore, all frequencies were represented.
Their average TA-EG score was 3.16 ($sd=0.54$). With a minimum of 1.85 and a maximum of 4.1 on a scale from 1 to 5.

Three additional participants were excluded. One quit the study after the first condition because the study was too time-consuming for that person. One had to stop during the \textsc{vr} condition due to pain in the eyes and one person could not finish the \textsc{vr} condition due to technical problems with the setup.

\subsection{Bystanders}
In total, we got 231 answers from bystanders. 
These answers were given by 209 different people. 
Twelve people answered questions multiple times. 
We only considered the first answer someone gave for each condition. This results in 69 answers for \textsc{laptop}, 82 answers for \textsc{ar} and 76 answers for \textsc{vr}.

Of these 209 people, 104 were female (50\%), 103 male (49\%), and 2 others (1\%). Their mean age was 23.99 ($sd=4.29$) and ranged from 18 to 45 with 22 people being at least 30 years old.
When asked about their occupation, 198 said they were students, 13 named a profession in which they are working, and 2 people said they are both studying and working. Large groups of students were studying in the areas of social work (60), business administration (37), computer science (30) and bioanalysis (24).

Bystanders rated their experience with each device on a scale from 1 (none) to 5 (very much). The mean score for \textsc{laptop} was 4.1 ($sd=4.17$), for \textsc{ar} 1.34 ($sd=1.49$) and for \textsc{vr} 1.8 ($sd=2.03$).
A Kruskal-Wallis-Test showed a significant main effect of \textsc{interface} on the experience ($H(2)=145.91, p<0.001, e^2=0.64$), and post-hoc tests using Dunn-Bonferroni-Tests indicated that bystanders were significantly more experienced in \textsc{laptop} than both \textsc{ar} ($p<0.001$) and \textsc{vr} ($p<0.001$) and also they were significantly more experienced in \textsc{vr} than \textsc{ar} ($p=0.03$).

Bystanders' technical enthusiasm on a scale from 1 (not at all) to 5 (very much) was on average 2.8 ($sd=1.08$). Their technical competency was on average 3.08 ($sd=0.8$), their positive attitude towards technology 3.44 ($sd=0.89$) and their negative attitude 2.75 ($sd=0.89$).

All bystanders said they have used a \textsc{laptop} in public before. Only 
13\% said this about \textsc{vr} devices and only 
10\% have used \textsc{ar} devices in public.
Locations in which bystanders have used \textsc{ar} or \textsc{vr} devices in public include universities or schools, museums, fairs, or game centers.
In contrast, 
78\% stated that they saw another person using a \textsc{laptop} in public before, 
33\% said they saw someone using a \textsc{ar} device in public before, and 
41\% saw someone using a \textsc{vr} device in public before.
However, there is reasonable doubt that some people did not understand the question correctly, as all people answered they had used a \textsc{laptop} themselves in public, but apparently only 78\% of them saw other people using one. When asked about which devices they used to work in public the most common answers were laptops (mentioned by 86\%), tablets (34\%) and smartphones (24\%). No one mentioned VR or AR devices.


\subsection{Methodology}
We used a repeated measures analysis of variance (RM-ANOVA) to analyze the data obtained through the study. The results can be found in Table \ref{tab:ANOVA}. For post-hoc comparisons, we applied Bonferroni adjustments at an initial significance level of $\alpha=0.05$. To ensure the robustness of the ANOVA, we used Greenhouse-Geisser correction in cases where the sphericity assumption was not met \cite{blanca2023non}.
For subjective data, we used aligned rank transform (ART) \cite{wobbrock2011aligned} and the ART-C method \cite{elkin2021aligned} for multiple comparisons, also using Bonferroni adjustments. 

We used two one-sided tests (TOST) to test the equivalence between conditions. The equivalence lower and upper bounds were calculated using Cohen's $d_z = \frac{t}{\sqrt{n}}$ with the critical $t$ value for $n=18,\alpha=0.05$ for two-tails (c.f., \cite{lakens2018equivalence}). The value of $d_z$ for our analysis was $\pm0.495$, with $-0.495$ as the lower bound and $+0.495$ as the upper bound. 
In the following we use $m$ to abbreviate the arithmetic mean and $sd$ for standard deviation.
We used axial coding to structure statements from free-text-fields and interviews.

\subsection{Results}

\begin{table}[t]
    \centering 
    \caption{RM-ANOVA results. }
    \small
    \setlength{\tabcolsep}{5pt}

    \begin{tabular}{|c|c|c|c|c|c|}
        \hline
          & $f_{1}$ & d$f_{2}$ & F & p & $\eta^2_p$     \\ 
        \hline
        threatened vs. safe & $2$ & $34$ & $16.824$ & $<0.001$ & $0.497$   \\
        \hline
        unsure vs. self-confident & $2$ & $34$ & $4.063$ & $0.026$ & $0.193$   \\
        \hline
        observed vs. unobserved & $2$ & $34$ & $6.214$ & $0.005$ & $0.268$   \\
        \hline
        System Usability & $2$ & $34$ & $5.117$ & $0.011$ & $0.231$   \\
        \hline
        TLX Physical Demand & $2$ & $34$ & $4.686$ & $0.016$ & $0.216$   \\
        \hline
        TLX Effort & $2$ & $34$ & $5.788$ & $0.007$ & $0.254$   \\
        \hline
        Visual Fatigue & $2$ & $34$ & $15.847$ & $<0.001$ & $0.482$   \\
        \hline
        Total Simulator Sickness & $2$ & $34$  & $9.609$ & $<0.001$ & $0.361$   \\
        
        \hline
        Number of Window Switching & $1.12$ & $13.5$& $16.0$ & $0.001$ & $0.572$ \\
        \hline
        
    \end{tabular}
    
    \label{tab:ANOVA}
\vspace{-0.6cm}
\end{table}

\begin{figure*}[t]
	\centering 
	\includegraphics[width=0.9\linewidth]{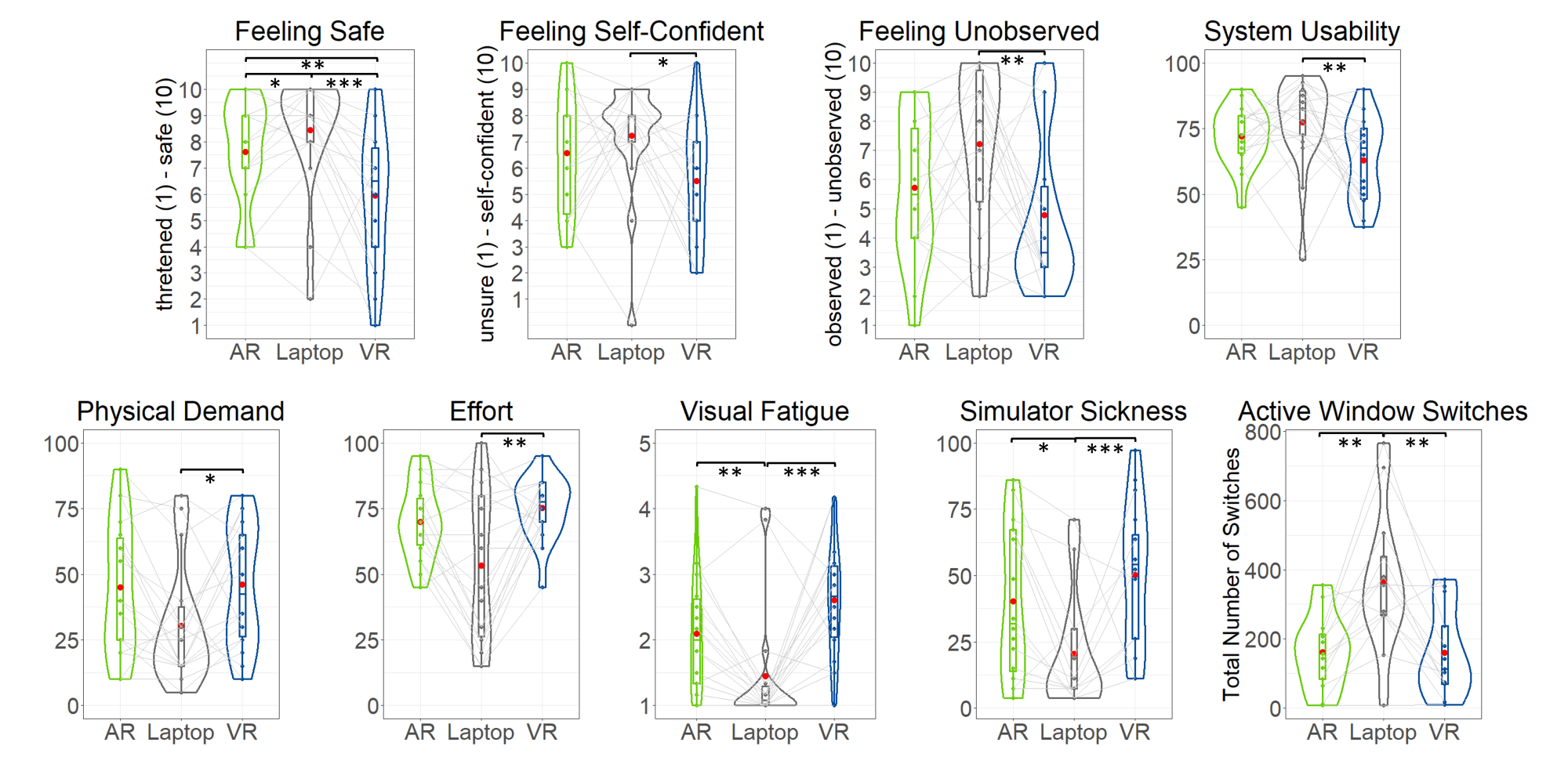}
	\caption{Violin plots for measures where significant differences were found. Red dots indicate the arithmetic mean and stars indicate the level of significance in post-hoc tests: $*<0.05$, $**<0.01$, $***<0.001$. }
	\label{fig:SignificantDifferences}
 \vspace{-0.5cm}
\end{figure*}

\paragraph{Being able to see the environment might increase the feeling of safety}
We found a significant main effect of \textsc{interface} on participants' rating of feeling more threatened or safe. Post-hoc tests indicate that they felt significantly more safe in \textsc{laptop} ($m=8.44,~sd=2.2$) as compared to both \textsc{AR} ($m=7.61,~sd=2.03$) ($p=0.04$) and \textsc{VR} ($m=5.94,~sd=2.67$) ($p=<0.001$) (see Fig. \ref{fig:SignificantDifferences}). In addition, participants felt significantly safer in \textsc{AR} than in \textsc{VR} ($p=0.009$) which was also mentioned by P12 in the interview.
Participants also stated that they liked that they could see the real environment in \textsc{ar} (P5, P15, P21, P22, P24) which made them feel more secure (P5) and in contrast, some participants (P16, P18, P24) didn't like not being aware of their physical environment in \textsc{vr}.
These combined findings might indicate that participants' feeling of safety is related to how well they can perceive their physical environment.
We did not observe any behavior that could have been dangerous for the participants.

\paragraph{XR can influence self-confidence}
We also found a significant main effect of \textsc{interface} on participants' rating of feeling more unsure or self-confident. Post-hoc tests indicate that they felt significantly more self-confident in \textsc{laptop} ($m=7.22,~sd=2.18$) as compared to \textsc{vr} ($m=5.5,~sd=2.38$) ($p=0.022$) (see Fig. \ref{fig:SignificantDifferences}).
In the \textsc{vr} condition, P18 reported to feel "like a show-off".
Even though we did not find significant differences between \textsc{ar} and the other conditions, some statements from participants indicate an impact on their self-confidence.
P16 did not like to stare at a screen placed in the direction of other people which relates to prior findings \cite{ng2021passenger, medeiros2022fromshielding}.
While walking to the cafeteria for the \textsc{ar} condition, which was her first, P7 mentioned she was nervous, and one bystander mentioned that the participant was brave for using \textsc{ar} in public.
These findings indicate that XR can have an influence on users self-confidence, especially in \textsc{vr}.

\paragraph{Using XR-HMDs currently makes you stand out}
We also found a significant main effect of \textsc{interface} on participants' rating of feeling more observed or unobserved. Post-hoc tests indicate that they felt significantly less observed in \textsc{laptop} ($m=7.22,~sd=2.76$) as compared to \textsc{VR} ($m=4.78,~sd=2.67$) ($p=0.004$) (see Fig. \ref{fig:SignificantDifferences}).
P23 also stated that he felt people were looking at him in \textsc{vr}.
From observing the bystanders, we know that this feeling is reasonable. 
We observed that, unsurprisingly, no bystander seemed to have any interest in the participants during the \textsc{laptop} condition, as they were blending in well with other people working on their laptops.
During the \textsc{ar} and \textsc{vr} conditions, individual bystanders seemed to be staring at the participants for multiple seconds, some also repeatedly. However, the majority of bystanders either did not, or pretended not to notice the participants or only looked at them in a very subtle way. Some seemed to find excuses to look such as when they walked past the participant.
In addition, while in the \textsc{laptop} condition only 65\% indicated that they noticed the participant, for \textsc{ar} 93\% and for \textsc{vr} even 95\% noticed the participant.
Fisher's exact test revealed that there is a significant effect of \textsc{interface} on whether bystander noticed the participant or not ($p<0.001$), with post-hoc tests showing that in the \textsc{laptop} condition they noticed the participant significantly less often than in \textsc{ar} ($p<0.001$) and \textsc{vr} ($p<0.001$).
These findings indicate that XR-users stand out, and that users feel that, especially in \textsc{vr}. However, the reactions of the bystanders remain mostly subtle.

\paragraph{Indications of XR benefits}
Even though some measures have shown to be in favor of using a \textsc{laptop}, participants value certain aspects of the XR-experiences.
This has been indicated by the participants choice of their overall most-liked system, in which 7 participants named \textsc{laptop}, 5 \textsc{ar} and 6 \textsc{vr} and therefore did not show a clear preference towards one.
Most participants did not like the small screen space in the \textsc{laptop} condition (P4, P6, P7, P9, P14, P15, P17, P18, P19, P21, P23, P24), found it annoying to switch between content (P12, P14, P22) and were distracted by the surrounding (P5, P8, P15).
Therefore, participants liked that they had more screen space in \textsc{ar} (P4, P6, P7, P9, P11, P12, P15, P17, P21, P23) which gave them a better overview of the information (P4, P7, P8, P14, P15, P18) and made them feel more productive in \textsc{ar} (P17)  
Similarly, participants liked the increased screen space in \textsc{vr} (P5, P6, P9, P17, P18, P19, P21), but also that they could be immersed in their own world (P7, P8, P22, P23) which could be customized (P16, P18, P19, P21, P23).
Participants also acknowledged the ergonomic benefits of \textsc{ar} such as the possibility to look straight ahead (not down at the laptop screen) (P15, P16), which of course also applies to \textsc{vr}.
Specifically for \textsc{vr}, participants mentioned they could concentrate better (P8, P15, P24), as they didn't see anything else which also made them feel more productive (P8, P24). 
But also, for \textsc{ar} P7 mentioned she noticed less from outside and P22 was less distracted than with the \textsc{laptop}.
In addition, through logging the number of active window switches (the user clicks in another window), we found a significant main effect of \textsc{interface}, even though, unfortunately, we had to exclude 5 participants due to logging errors. 
Post-hoc tests showed, that in \textsc{laptop} ($m=366.62,~sd=204.76$) participants switched the active window significantly more often than in both \textsc{ar} ($m=161.54,~sd=107.08$) ($p=0.005$) and \textsc{vr} ($m=159.69,~sd=126.4$) ($p=0.004$) (see Fig. \ref{fig:SignificantDifferences}).
These observations support the findings of prior work that XR can provide certain benefits to knowledge workers. 

\begin{figure*}[t]
	\centering 
	\includegraphics[width=0.9\linewidth]{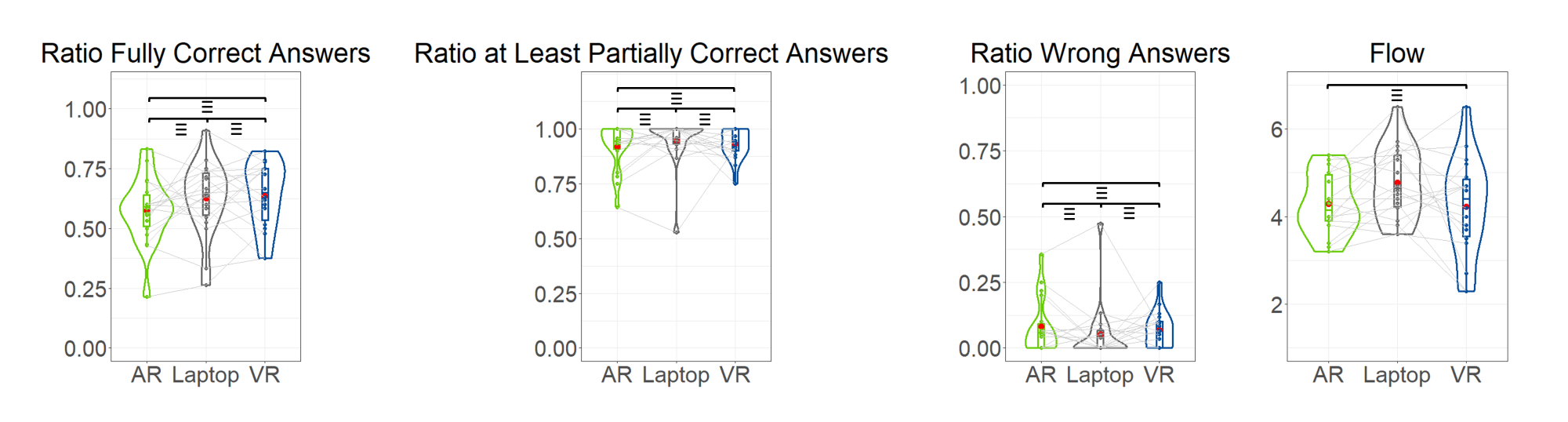}
	\caption{Violin plots for measures where equivalences were found. Red dots indicate the arithmetic mean and equivalences are indicated by $\equiv$. }
	\label{fig:Equivalences}
 \vspace{-0.5cm}
\end{figure*}

\paragraph{Similar diligence in all conditions}
We found an equivalence effect of \textsc{interface} on the ratios of fully correct answers between any two pairs of \textsc{laptop} ($m=0.62$, $sd=0.16$), \textsc{AR} ($m=0.58$, $sd=0.14$), and \textsc{VR} ($m=0.63$, $sd=0.14$), with both the two p-values being $p<0.001$ in all pairs. 
We also found an equivalence effect of \textsc{interface} on the ratios of at least partially correct answers between any two pairs of \textsc{laptop} ($m=0.94$, $sd=0.11$), \textsc{AR} ($m=0.92$, $sd=0.11$), and \textsc{VR} ($m=0.93$, $sd=0.07$), with both the two p-values being $p<0.001$ in all pairs. 
Similarly, the equivalence effect of \textsc{interface} was also found on the ratios of wrong answers between any two pairs of \textsc{laptop} ($m=0.05$, $sd=0.11$), \textsc{AR} ($m=0.08$, $sd=0.11$), and \textsc{VR} ($m=0.07$, $sd=0.07$), with both the two p-values being $p<0.001$ in all pairs (see Fig. \ref{fig:Equivalences}).
These findings show a similar level of diligence between all conditions, indicating that this was not influenced by the \textsc{interface}, neither positively nor negatively.
In addition, we found an equivalence effect of \textsc{interface} on flow between \textsc{AR} ($m=2.72$, $sd=0.57$) and \textsc{VR} ($m=2.75$, $sd=0.47$), with the larger of the two p-values being $p=0.042$ (see Fig. \ref{fig:Equivalences}).
This indicates that even though participants were more isolated in \textsc{vr} this did not influence their perceived level of flow compared to \textsc{ar}.

\paragraph{Usability of \textsc{laptop} higher than that of \textsc{vr}}
We found a significant main effect of \textsc{interface} on usability. Post-hoc tests showed that \textsc{laptop} ($m=77.36,~sd=17.11$) resulted in a significantly higher usability than \textsc{vr} ($m=62.91,~sd=15.81$) ($p=0.009$) (see Fig. \ref{fig:SignificantDifferences}).
Also, a significant main effect of \textsc{interface} was found on effort, with \textsc{laptop} ($m=53.33,~sd=29.0$) resulting in a significantly lower effort than \textsc{vr} ($m=75.28,~sd=11.82$) ($p=0.005$) (see Fig. \ref{fig:SignificantDifferences}).
In the interviews, participants mentioned that they liked the familiarity (P6, P7, P8, P9, P14, P15, P16, P21, P22, P23) of the \textsc{laptop} condition, that they could see the keyboard (P8, P17) and type easier (P14) and that everything was compact (P7, P11).
In contrast, some participants found \textsc{vr} slightly blurry (P6, P17), especially the pass-through (P18) so that they could not clearly see the keyboard (P18, P19). 
Usability issues reported for \textsc{ar} included losing the cursor due to the small field of view (P5, P9, P24), and the need to rotate the head to see everything (P7). Participants also mentioned the images were not always perfectly sharp (P7, P14, P23,), some noticed a flickering (P8, P14, P17) and they found the motion blur when moving the head especially annoying (P9, P16, P19, P23). Also, some mentioned that a bright environment made it hard to see the virtual content (P17, P18, P22) and P21 found it distracting to see the physical screen behind the virtual screens.
However, participants liked that they could see the keyboard underneath the HMD (P8, P15, P24) in the \textsc{ar} condition, which could contribute to the slightly, yet not significantly, higher usability score, compared to \textsc{vr}.
It is noteworthy, however, that the usability scores between participants vary quite a bit, such that the group of participants who overall liked \textsc{vr} best, rated the usability of \textsc{vr} ($m=77.92,~sd=7.32$) significantly higher ($p=0.002$) than the group who preferred \textsc{laptop} or \textsc{ar} ($m=55.42,~sd=13.35$).
Similarly, P19 and P23 said they preferred \textsc{vr}, because it was easy to use while P8 found it hard to remember all settings in \textsc{vr}.
These findings point out several usability problems of both \textsc{ar} and \textsc{vr}, while only \textsc{vr} has been rated significantly worse than the baseline \textsc{laptop} condition.

\paragraph{Current XR imposes physical discomfort}
We found a significant main effect of \textsc{interface} on physical demand and post-hoc tests showed that \textsc{laptop} ($m=30.28,~sd=22.19$) resulted in a significantly lower physical demand than \textsc{vr} ($m=46.11,~sd=23.36$) ($p=0.025$) (see Fig. \ref{fig:SignificantDifferences}).
Also, we found a significant main effect of \textsc{interface} on the visual fatigue with post-hoc tests showing, that it was significantly lower for \textsc{laptop} ($m=1.44,~sd=0.92$) compared to both \textsc{AR} ($m=2.09,~sd=0.9$) ($p=0.003$) and \textsc{VR} ($m=2.6,~sd=0.88$) ($p=<0.001$) (see Fig. \ref{fig:SignificantDifferences}).
In addition, we found a significant main effect of \textsc{interface} on the total simulator sickness score. Again, post-hoc tests showed that it was significantly lower for \textsc{laptop} ($m=22.64,~sd=36.13$) compared to both \textsc{AR} ($m=45.71,~sd=44.06$) ($p=0.025$) and \textsc{VR} ($m=56.51,~sd=31.58$) ($p=<0.001$) (see Fig. \ref{fig:SignificantDifferences}).
This also matches the answers from participants when asked in which condition they felt most comfortable, to which 16 participants answered \textsc{laptop}, 2 \textsc{ar} and no one \textsc{vr}. Fisher's exact test showed a significant association between \textsc{interface} and the choice of the most comfortable device ($p<0.001$) and post-hoc tests showed that participants were choosing \textsc{laptop} significantly more often as most comfortable than \textsc{ar} ($p<0.001$) or \textsc{vr} ($p<0.001$).
Participants preferred \textsc{laptop}, because they did not need to wear something (P9, P23), because no head movement was needed (P5, P9), and because it was most familiar (P8, P15) and normal (P17, P19).
P14 said he could work in \textsc{laptop} without feeling tired and P5 liked that there was no weight on her head.
Even though the Lenovo Think Reality glasses are relatively light, participants mentioned it pressed against their head (P12, P23, P24), were heavy on the nose (P15), caused eye strain (P4, P14) and a headache (P4).
Some concluded that they could only work in \textsc{ar} for a short time (P4, P12, P23), however P4 and P12 would actually prefer it over \textsc{laptop} for a short duration.
We also observed P5 supporting the \textsc{ar}-HMD a lot, P24 took a break for some minutes before continuing with the second task and P19 looked troubled from the glasses.
Similar answers were given with regards to \textsc{vr}, where some people had headaches (P4, P7, P11, P21), eye-problems (P4, P11, P12, P16, P17), and felt slightly sick (P7) or dizzy (P15).
They also mentioned that the \textsc{vr}-HMD pressed against their head uncomfortably (P8, P14, P21, P22) and was heavy (P15, P23) or they simply didn't like to have something on their head (P5, P9).
Again, some participants reported they wouldn't like to use it for long (P7, P17) and felt like they got tired faster (P12).
During \textsc{vr} conditions, we also observed that P4 was again repeatedly massaging his forehead, some adjusted the HMD (P7, P8, P11) or supported it for some time (P8, P15) and P21 seemed to have some pain.
These findings show that XR still imposes physical discomfort on the users, which seems to be mainly caused by the hardware specifications.

\paragraph{Personality and experience can influence the perception of XR-use in public}
We found that the personality trait “extraversion” negatively correlates with the overall-impression in AR ($\rho(16)=-0.6,~p=0.009$). This indicates that more extroverted participants found the \textsc{ar} experience more awkward.
In addition, we found a negative correlation between AR-experience and the feeling of being unobserved in the AR condition ($\rho(16)=-0.54,~p=0.022$) and analogously we found a negative correlation between VR-experience and the feeling of being unobserved in VR ($\rho(16)=-0.5,~p=0.037$)
This indicates that people with more \textsc{ar} or \textsc{vr} experience felt more observed in the corresponding condition.
On the other hand, the feelings of bystanders in the \textsc{vr} condition were found to positively correlate with their \textsc{vr}-experience ($r(60)=0.26,p=0.033$) indicating that with increased \textsc{vr}-experience they felt less foolish watching the participant.
These findings show that both personality and prior experience can have an influence on how users and also bystanders perceive XR-use in public.

\paragraph{Bystanders mostly feel confused or curious}
In a free-text-field, bystanders were also invited to describe their feelings in more detail. Not all bystanders did this, and some expressed multiple thoughts and feelings, sometimes contradictory. 
For the \textsc{laptop} condition, 30 statements were neutral, saying the behavior of the participant is normal, or they do not care.
In contrast, for the \textsc{ar} condition only 13 statements were neutral, 15 expressed feelings like confusion or uncertainty, also 15 mentioned it was cool and exciting and another 15 found it interesting and were curious about the device. In addition, 10 were concerned with questions such as “What is this?”, “How does it work?” and “What is the person doing?”. Some bystanders expressed the wish to also try the \textsc{ar} device (4) or were wondering if they were being observed (6).
After observing the \textsc{vr} condition, only 12 statements were neutral, 23 of them expressed feelings like confusion or unfamiliarity, 13 expressed that it was exciting and amazing and 9 mentioned it was interesting. Again, people were asking questions about the device and what the participant was doing (9) or if they could see them (3). Some again expressed the wish to try it (3).
Three bystanders explicitly mentioned they would not use \textsc{vr} in public.
One bystander said the participant looked foolish wearing the \textsc{vr} HMD, another said it was careless and two bystanders insulted the participants, calling them a victim and fool.

There were also several notable reactions. 
During \textsc{vr} conditions, 5 people looked at the participant and laughed (P4), two friends of P6 laughed at her and asked her if she could see them, two people were looking at P15 and giggling while talking to each other and a group of people looked at and talked about P16 and pointed him out to other friends.
During P5's \textsc{VR} condition one bystander got up, saw the participant, jumped back to his friend very excited, laughed and said “Is this the Quest Pro?”. When passing P5 again he even waves at her. Two bystanders stared at P8 for a while and discussed if they could ask if they could try it.
In \textsc{ar} three people sat down next to P15, as they approached they said “Ohaa” and one of them waved at P15.
Also, in \textsc{ar} P15, P19, P21 and P23 were talked to for a short time.
Overall, the majority of comments were either neutral, expressing positive feelings such as curiosity, or confusion with the situation.
Also, the bystanders who were more obvious with their reactions seemed rather interested than hostile, and negative comments were scarce.

Moreover, some bystanders who visited the cafeteria frequently were initially intrigued by the \textsc{ar} and \textsc{vr} headsets and observed the participants. But, over time, they did not seem curious or stare at any participants wearing those headsets, in their later visits. This could be an indication of rising acceptance of these headsets in public if people are exposed to them frequently - similar to the technology acceptance process of prior technologies \cite{edwards2003wearable, kim2009investigating}.

\paragraph{Bystanders understanding of participant actions is less clear in XR}
We also included a text-field in which bystanders could speculate about what they thought the participant was doing.
For the \textsc{laptop} condition the most common answers were something related to studying at the university (39) or working (13).
In the \textsc{ar} condition only 19 bystanders thought the participant was doing something study-related and only 10 thought about working. However, 13 though the participant is playing a game. Eight were thinking about virtual worlds or content, 5 about an experiment, 4 about development and testing, and 4 thought the participant was busy with emails, videos, or news.
Four people also speculated that the participant might have eye-problems. 
Regarding the \textsc{vr} condition, 25 bystanders thought the participant was doing something study-related and 12 thought about work. This time, 20 bystanders even thought the participant was playing games, 9 said they were testing something and 4 thought about programming. Also, 6 bystanders thought the participants might be designing or constructing something. Two said they are in a virtual world, 3 said they are doing daily tasks such as videos and browsing the internet and 3 even mentioned the participant is using a virtual office with one bystander even mentioning an increased virtual screen.
These observations show that while bystanders were generally thinking that the participant was studying or working when using the \textsc{laptop}, they were more often thinking about gaming when seeing \textsc{ar} and \textsc{vr}. Also, there was a wider range of guessed activities than in \textsc{laptop}.

\paragraph{Isolation from physical surroundings}
Sometimes, bystanders were sitting down very closely to the participants. Four participants (P4, P6, P14, P16) experienced this in \textsc{ar} but reported that they did not care. P16 was just slightly distracted when a bystander sitting close by laughed.
In \textsc{vr}, P7 noticed someone sat down next to her, but was wondering how close and P12 didn't even notice someone sitting very closely. 
Through the passthrough, P5 noticed some people staring at her, but she did not see the bystander who was waving at her.
P19 and P21 were talked to in \textsc{ar} and reported they felt normal about it, but P21 was concerned about getting back to the task as he wanted to perform well in the study.
P24 was touched by a friend during the \textsc{vr} condition and was surprised at first, but quickly figured out who it must be. This is plausible as the study took place in the university cafeteria. In a completely unfamiliar environment, the participant might have reacted differently. 

Only P5 used full passthrough throughout the \textsc{vr} condition, and P9 switched to passthrough occasionally. All other participants decided to use one of the virtual environments, mostly a mountain-lodge or space-themed environments, in which they felt comfortable or which they thought looked nice.
Of the ones using a virtual environment, 14 participants did not use a passthrough window as they thought it was distracting (P4, P14, P16), was not needed (P7, P11, P17, P18, P19, P23), felt more focused and immersed (P23, P24) or felt like they had less space for the task (P12). However, P6 regretted it, as she wanted to see who was talking.
Of the three participants who used a passthrough window, P8 moved it out of sight as it interfered with the task, and P9 and P22 had one on both sides but didn't really use it during the task.
From these limited observations, it seems like participants noticed less of their environment in \textsc{vr} than \textsc{ar} which was also reported by \cite{li2022designing}. This might also be influenced by the fact that most participants did not use passthrough windows or did not actively use them.

\paragraph{No significant influence on productivity}
In the final interview, participants were also asked in which condition they felt most productive. There, 11 participants felt most productive in \textsc{laptop}, 2 in \textsc{ar} and 5 in \textsc{vr}.
However, we could not find a significant main effect of \textsc{interface} regarding the total given answers, the number of fully correct, partially correct or wrong answers.
We also did not find significant correlations between participants productivity and their experience with multi-monitor setups or \textsc{ar} and \textsc{vr}.
Some participants mentioned they felt most productive in \textsc{laptop} because it was more familiar (P11, P15, P22) and did not make them tired (P14).
However, P11 and P22 mentioned they could be more productive in \textsc{ar} or \textsc{vr} if they used it more often.
P17 felt most productive in \textsc{ar} because of the large screen space and P8 felt more productive in \textsc{vr} because she could concentrate better.

\paragraph{Preconditions for using XR in future}
Analogous to \cite{ahlstrom2014you}, we asked participants where and in front of whom they would feel comfortable using the respective devices.
Fourteen participants would feel comfortable using a \textsc{laptop} everywhere, some named limitations such as not outside (P8, P17), needing a quiet place (P19, P21) or not a location that is dynamically changing (P22). All participants said they would use the \textsc{laptop} in front of anyone.
Ten participants would use \textsc{ar} anywhere, as it is low-key (P9), they can see the environment (P14) and don't mind what others think as they don't see them (P6).
Six participants would prefer more private places, as they might feel observed (P7). 
Similar to the \textsc{laptop}, some prefer indoors (P17, P18), quiet rooms (P18, P21) and less dynamic (P22) or crowded places (P23).
Eleven participants would use \textsc{ar} in front of anybody, with exceptions such as feeling safe (P22, P23). 
Others prefer people who know of the device or maybe also use it (P7, P8, P14, P15, P16, P24) and P12 and P24 would prefer people they know.
Only six participants would use \textsc{vr} anywhere, as they can not see the others (P4). 
Others would rather use it in only semi-public spaces such as a university (P9, P15), or in environments that provide some privacy such as a capsuled seat (P8, P16, P21). 
Some would rather use it in spaces like offices or when being alone (P6, P12, P16, P17, P18) or in stationary places (P5, P22). 
Others were concerned with being unsocial (P8) or about people's reactions (P6, P17). 
Some would prefer silent spaces (P21) or ones that are not too tight (P23).
Ten participants would use \textsc{vr} in front of anybody, under certain conditions such as feeling safe (P5). Some people would only use it in the company of people they know (P8, P12, P16, P22) or people who know about the \textsc{vr} device (P14, P18, P21).
For both \textsc{ar} and \textsc{vr}, P6 would prefer to use them in front of strangers, as friends might make fun of her.

When asked, if they could imagine using the device in their future, five people agreed for \textsc{ar}, while the other 13 could imagine it, some under preconditions such as only for a short time (P4, P14), if the field of view was larger (P5), if it was lighter (P22) and fits better with normal glasses (P18) or if it was cheap enough (P17). Some participants especially mentioned the use case of not having a fixed physical setup such as while traveling (P6, P9, P11, P15, P17, P22).
Twelve participants would like to use \textsc{vr} in the future, some of them stating some conditions such as it being more optimized (P19), only for short times (P4, P5), if it offers more advantages such as making use of the 3D display (P11), being more lightweight (P22) or being available for free (P14). Again, they mention to use it on the go (P7, P9) or to exchange reality for a better environment (P8). 
Four participants would not like to use \textsc{vr} in the future, and another two would give it another try before deciding.

These observations indicate, that for both \textsc{ar} and \textsc{vr} many participants name certain requirements for places in which they would use these devices, such as semi-private rooms or less crowded or dynamic spaces. 
However, while all participants could imagine using \textsc{ar}, at least under some preconditions, 4 people could not see themselves using \textsc{vr} in the future. For both devices, participants would like to see some improvements first, mainly in hardware, such as better resolution and less weight.

\paragraph{Subtle differences between AR and VR}
As reported before, participants felt significantly safer in \textsc{ar} than \textsc{vr}.
Apart from that, participants reported that \textsc{ar} felt good (P11), was not that heavy (P21, P11) and was simple to set up (P19).
On the other hand, participants valued \textsc{vr} because there was no motion blur (P19), the screens were better (P7) and they could see more (P9).
Also, P9 found that the weight was better balanced than in \textsc{ar} and P24 was less distracted by seeing other people. 
This could indicate that \textsc{ar} can score with a higher feeling of safety and more comfort, while \textsc{vr} provides a better visual experience with fewer distractions. 
Generally, we observed that there was a great difference between participants with some strongly preferring \textsc{ar} and some \textsc{vr}.

\section{Discussion and Limitations}
Our quantitative measures demonstrated several significant differences between \textsc{laptop} and XR, and our observations in combination with participants' and bystanders' statements gave us a more nuanced understanding of how they perceived the use of XR in public.

There were indications that participants felt safer the more they could see their physical surroundings. 
The lack of being able to see the physical world clearly also affected the usability of \textsc{vr} as some participants raised concerns about not being able to clearly see the keyboard.
On the other hand, seeing other people can also restrict the placement of virtual screens, as was also shown in prior work \cite{ng2021passenger, medeiros2022fromshielding}.
Most participants reported that they did not use passthrough windows in \textsc{vr} as they considered it unnecessary, or even distracting. This calls for more research on how to better integrate passthrough into \textsc{vr} so that users can benefit from seeing their surroundings without distracting them or limiting the advantages of \textsc{vr}.

We also found that bystanders noticed XR users more frequently than \textsc{laptop} users. Similarly, XR users feel more observed, especially in \textsc{vr}.
However, this could change in the future, if XR becomes more commonplace and more people are using it.
More frequent use of XR would also increase the familiarity with the technology, which was currently found the be one of the main advantages of \textsc{laptop}. Two participants even mentioned that they think they would be more productive with XR---if they used it more often.
We also observed that many bystanders were curious about the devices or even expressed the wish to try them. This could indicate that people can become open to using such systems. 
However, bystanders believed participants were studying or working in XR less often than for \textsc{laptop}. This might be because current XR technologies, especially \textsc{vr}, is widely known for gaming and less as a tool for knowledge work. Yet, this could already be in the process of changing, as some current XR products are advertised specifically as a tool for knowledge work.

Our findings indicate that there are still some required hardware improvements for XR to be viable for prolonged knowledge work. In particular, there is a need for a larger field of view in \textsc{ar}, and for both \textsc{ar} and \textsc{vr} the display quality and form factor should be improved. Currently, several participants would only use the XR-devices for a short time, even though they can see benefits, such as increased screen space and fewer distractions.

We also found that personal preferences likely to play an important role in whether participants preferred \textsc{ar} or \textsc{vr}, as some were very enthusiastic about the devices while others expressed a strong dislike for them.
In addition, there were indications that personality and prior XR experience can influence feelings while using XR in public, or watching others doing so. Yet, we only collected a small range of data and further research could uncover more factors that could for example explain preferences for certain XR-devices.

Some participants reported that they might not use XR in widely open public spaces. They proposed some semi-private spaces such as offices or libraries or places that are less dynamic and in which participants have a dedicated, maybe partially encapsulated, space, such as a seat in a train or an airplane.
Both XR-devices seem to have certain advantages, such as being more aware of the surroundings in \textsc{ar} and having a slightly better screen in \textsc{vr}. Therefore, we could not find a general preference in favor of \textsc{ar} or \textsc{vr}. Choosing a device could therefore also depend on the situation and use-case. XR-devices combining both technologies could allow users to making a choice depending on their needs.

We decided to use a university cafeteria for the study. Even though this cafeteria was open to everyone, it was still a rather exclusive setting with most people belonging to the community of university students. This could have an influence on their reactions towards their peers participating in the study. Therefore, similar studies should be repeated in different public and semi-public settings.
Also, the study itself could have an influence on participants' feelings, such that they might feel more confident because they are contributing to a study and not choosing to use XR-devices on their own accord.
In addition, the results on productivity can heavily depend on the task, and results can therefore vary for other types of work.

\section{Conclusions}
With the goal of observing how using XR devices in public affects the users and also the bystanders, we conducted a study in which participants used an \textsc{ar} and \textsc{vr} HMD as well as a \textsc{laptop} in a university cafeteria.
Our results indicate that using XR can influence the self-confidence, that currently using XR in public still makes users stand out and that users' personality and prior XR experience can influence how they feel about using XR in public.
We also observed that bystanders were often curious about the XR users but had no clear understanding of what they were doing. This might change in the future, as XR becomes more common, also as a tool for knowledge work.
In addition, we found indications that seeing the virtual surrounding can increase users feeling of safety, however, most participants said passthrough windows in \textsc{vr} were unnecessary or distracting. Therefore, more research is required to integrate passthrough in a sensible way.
Finally, even though participants acknowledge the benefits of XR, the tested XR devices still have lower usability and comfort compared to the baseline \textsc{laptop}, and, therefore, further improvements of their hardware and software are needed to establish them as a tool for knowledge work that can also be used in public.


\bibliographystyle{abbrv-doi}
\balance

\bibliography{template}
\end{document}